\documentclass[superscriptaddress,twocolumn,amsmath,amssymb,floatfix,prb,a4paper]{revtex4}

\usepackage{graphicx}
\usepackage{dcolumn}
\usepackage{bm}
\usepackage{color}
\usepackage{units}
\usepackage{booktabs}
\usepackage{multirow}
\usepackage{tabularx}

\newcommand{\superscript}[1]{\ensuremath{^\textrm{#1}}}
\newcommand{\subscript}[1]{\ensuremath{_\textrm{#1}}}

\newcommand{\general}{Rb$_{x}$Mn[Fe(CN)$_6$]$_{\frac{(2+x)}{3}}$$\cdot$zH$_2$O}
\newcommand{\B}{Rb$_{0.81}$Mn[Fe(CN)$_6$]$_{0.95}$$\cdot$1.24H$_2$O}
\newcommand{\A}{Rb$_{0.97}$Mn[Fe(CN)$_6$]$_{0.98}$$\cdot$1.03H$_2$O}
\newcommand{\C}{Rb$_{0.70}$Cu$_{0.22}$Mn$_{0.78}$[Fe(CN)$_6$]$_{0.86}$$\cdot$2.05H$_2$O}
\newcommand{\cn}{$\nu_{\textrm{CN}}$}
\newcommand{\figwidth}{7.5cm}

\begin{document}

\title{Bulk and surface switching in Mn-Fe-based Prussian Blue Analogues}

\author{T.T.A. Lummen}
\author{R.Y.N. Gengler}
\author{P. Rudolf}
\author{F. Lusitani}
\affiliation{Zernike Institute for Advanced Materials, University of
Groningen, Nijenborgh 4, 9747 AG Groningen, The Netherlands}
\author{E.J.M. Vertelman$^{1,}$}
\author{P.J. van Koningsbruggen$^{1,}$}
\affiliation{Stratingh Institute for Chemistry, University of
Groningen, Nijenborgh 4, 9747 AG Groningen, The Netherlands}
\author{M. Knupfer}
\author{O. Molodtsova}
\affiliation{IFW Dresden, P.O. Box 270116, D-01171 Dresden, Germany}
\author{J.-J. Pireaux}
\affiliation{L.I.S.E., Facult\'{e}s Universitaires Notre-Dame de la
Paix, Rue de Bruxelles 61, B-5000 Namur, Belgium}
\author{P.H.M. van Loosdrecht}
\email{P.H.M.van.Loosdrecht@rug.nl} \affiliation{Zernike Institute
for Advanced Materials, University of Groningen, Nijenborgh 4, 9747
AG Groningen, The Netherlands}

\begin{abstract}
Many Prussian Blue Analogues are known to show a thermally induced
phase transition close to room temperature and a reversible,
photo-induced phase transition at low temperatures. This work
reports on magnetic measurements, X-ray photoemission and Raman
spectroscopy on a particular class of these molecular
heterobimetallic systems, specifically on \A, \B\ and \C, to
investigate these transition phenomena both in the bulk of the
material and at the sample surface. Results indicate a high degree
of charge transfer in the bulk, while a substantially reduced
conversion is found at the sample surface, even in case of a near
perfect (Rb:Mn:Fe=1:1:1) stoichiometry. Thus, the \textit{intrinsic}
incompleteness of the charge transfer transition in these materials
is found to be primarily due to surface reconstruction. Substitution
of a large fraction of charge transfer active Mn ions by charge
transfer inactive Cu ions leads to a proportional conversion
reduction with respect to the maximum conversion that is still
stoichiometrically possible and shows the charge transfer capability
of metal centers to be quite robust upon inclusion of a neighboring
impurity. Additionally, a 532 nm photo-induced metastable state,
reminiscent of the high temperature
Fe\superscript{III}Mn\superscript{II} ground state, is found at
temperatures 50-100 K. The efficiency of photo-excitation to the
metastable state is found to be maximized 90 K. The photo-induced
state is observed to relax to the low temperature
Fe\superscript{II}Mn\superscript{III} ground state at a temperature
of approximately 123 K.
\end{abstract}

\date{\today}

 \maketitle

\textbf{Keywords:} Prussian Blue Analogues, Raman spectroscopy, XPS,
Magnetic properties, charge transfer transitions, photo-induced
transitions.

\section{Introduction}
In recent history, a substantial amount of scientific interest has
been directed towards photomagnetic materials from a technological,
application-oriented point of view, due to their favorable
properties.\cite{sato04} Among materials displaying photo-induced
spin-crossover\cite{decurtins84,decurtins85,hauser99,guttlich00,boillot04}
and valence
tautomerism\cite{carducci97,caneschi98,hendrickson04,carbonera04,carb04},
an important subclass is formed by the so-called Prussian Blue
Analogues. These molecular heterobimetallic coordination compounds
exhibit intervalence charge transfer transitions induced by various
external stimuli (temperature\cite{ohk05},
pressure\cite{ksenofontov03,morimoto03}, visible
light\cite{sato04,ohk05} and X-rays.\cite{margadonna04}) A prominent
member of this subclass is \general\
\cite{tokoro03,morimoto03,ohkoshi05,kat03}, which undergoes a
temperature-induced charge transfer transition from a high
temperature cubic lattice, space group $F$\={4}$3m$ (HT phase), to a
low temperature tetragonal ($I$\={4}$m2$) phase (LT phase). This
reversible, entropy-driven\cite{luzon08,cobo07} phase transition,
which is described by Fe\superscript{III}($t^{5}_{2g}$,
S=$\nicefrac{1}{2}$)-CN-Mn\superscript{II}($t^{3}_{2g}e^{2}_{g}$,
S=$\nicefrac{5}{2}$) \textbf{$\rightleftharpoons$}
Fe\superscript{II}($t^{6}_{2g}$,
S=$0$)-CN-Mn\superscript{III}($t^{3}_{2g}e^{1}_{g}$, S=$2$), occurs
not only under the influence of temperature (HT$\rightarrow$LT at
$\sim$ 225 K, LT$\rightarrow$HT at $\sim$ 290 K), but can also be
induced by visible light irradiation at various
temperatures\cite{tokoro03,tokoro05,mo03,cobo07,tokoro08}, by
hydrostatic pressure\cite{morimoto03} and possibly by X-ray
radiation.\cite{margadonna04} In addition, these type of Prussian
Blue Analogues have demonstrated a variety of other interesting
properties such as a pressure-induced magnetic pole
inversion\cite{egan06} and multiferroicity\cite{ohkoshi07}. The
capability of these materials to display switching phenomena,
however, is known to depend rather crucially on its exact
stoichiometry.\cite{ver06,cobo07,ohkoshi05} Even though it has been
established that the degree of conversion in these materials is
maximized for systems closest to a Rb:Mn:Fe stoichiometry of 1:1:1,
there appears to be an intrinsic limit to the maximum conversion
achieved. To our knowledge, no \general\ system has ever been shown
to undergo a complete transition from either configuration
(Fe\superscript{III}Mn\superscript{II} or
Fe\superscript{II}Mn\superscript{III}) to the other. That is, all
data on \general\ systems seem to indicate the presence of at least
small amounts of the HT configuration
(Fe\superscript{III}Mn\superscript{II}) when in the LT phase
(configuration Fe\superscript{III}Mn\superscript{II}) and often also
vice versa. The present understanding is that this incompleteness
originates from the intrinsic local inhomogeneities of these
materials, such as Fe(CN)$_{6}$-vacancies and alkali ion
nonstoichiometry. This paper investigates where exactly the
intrinsic incompleteness of the transition stems from in two ways:
Firstly, by comparing the charge transfer (CT) properties of the
bulk material to those of the surface material for both a 'near
perfect' (close to 1:1:1 stoichiometry) sample and a less
stoichiometric sample. Secondly, by quantitatively investigating the
effect of substituting part of the metal ions involved in the CT
transition by CT-inactive ions. Three different samples, \A, \B, and
Rb$_{0.70}$Cu$_{0.22}-$
Mn$_{0.78}$[Fe(CN)$_6$]$_{0.86}$$\cdot$2.05H$_2$O are quantitatively
compared utilizing different experimental techniques, which are all
capable of distinguishing between the two configurations
(Fe\superscript{III}Mn\superscript{II} (HT) and
Fe\superscript{II}Mn\superscript{III} (LT)). Magnetic measurements
are performed to obtain information on the bulk properties of the
various samples, while XPS spectroscopy is used to extract the
surface properties. Finally, Raman scattering is employed as a
tertiary probe to investigate the materials' properties.

\section{Experimental Methods}
\subsection{Sample synthesis} All chemicals (of analytical grade) were purchased at Sigma-Aldrich and used without further purification.
\A\ (sample A) and \B\ (sample B) are, respectively, samples 3 and 4
of a previous publication.\cite{ver06} Their synthesis and detailed
initial characterization can be found there. \C\ (sample C) was
prepared similarly, by slowly adding a mixed aqueous solution (25mL)
containing CuCl$_{2}$$\cdot$$2$H$_{2}$O (0.085 g, 0.02 M) and
MnCl$_{2}$$\cdot$$4$H$_{2}$O (0.396 g, 0.08 M) to a mixed aqueous
solution (25 mL) containing K$_{3}$[Fe(CN)$_{6}$] (0.823 g, 0.1 M)
and RbCl (3.023 g, 1 M). The addition time was 20 minutes and the
resulting solution was stirred mechanically and kept at a
temperature of 50$^{\circ}$C both during the addition time and for
the subsequent hour. A brown powder precipitated. This was
centrifuged and washed twice with distilled water of room
temperature. The powder was allowed to dry in air for about 12 hours
at room temperature. Yield (based on Mn + Cu): 81 \%. Elemental
analysis (details in Supporting Information) showed that the
composition of sample C was
Rb$_{0.70}$Cu$_{0.22}$Mn$_{0.78}$[Fe(CN)$_{6}$]$_{0.86}$ $\cdot$2.05
H$_{2}$O. X-ray powder diffraction showed that the sample was
primarily (weight fraction 80.4(8) \%) in the typical $F$\={4}$3m$
phase with the other fraction (19.6(8) \%) in the $I$\={4}$m2$
phase. The sample was confirmed to be single phase; phase separation
into \general\ and
Rb$_{x}$Cu[Fe(CN)$_6$]$_{\frac{(2+x)}{3}}$$\cdot$zH$_2$O fractions
was excluded. See for details the Supporting Information. As noted
in a previous paper\cite{ver06}, the samples under discussion here
deviate from a perfect Rb:Mn(+Cu):Fe stoichiometry of 1:1:1. This
deviation is ascribed to [Fe(CN)$_{6}$]$^{3-}$
vacancies\cite{ver06,ohkoshi05,cobo07} which are filled by H$_{2}$O
molecules, consistent with the hydration found in the materials.

\subsection{Instrumentation and measurement}
\emph{Magnetic measurements.} Magnetic measurements were performed
on a Quantum Design MPMS magnetometer equipped with a
superconducting quantum interference device (SQUID). Samples were
prepared by fixing 20-30 mg of the compound (0.5 mg weight accuracy)
between two pieces of cotton wool in a gelcap. For the magnetic
susceptibility measurements of samples A and B, the samples were
first slowly cooled from room temperature to 5 K (to ensure the
samples are not quenched in their HT phase\cite{tokoro06}). Then the
field was kept constant at 0.1 T while the temperature was varied
from 5 K to 350 K and back to 150 K (rate $\leq$ 4 K/min.). For the
magnetic susceptibility measurements of sample C the field was kept
constant at 0.1 T while the temperature slowly varied from 330 to 5
K and back to 330 K.

\emph{X-ray photoemission spectroscopy.} X-ray photoemission
spectroscopy (XPS) data were collected at the IFW Leibniz Institute
for Solid State and Materials Research in Dresden, using a SPECS
PHOIBOS-150 spectrometer equipped with a monochromatic Al
K$_{\alpha}$ X-ray source ($h\nu = 1486.6$ eV); the photoelectron
take off angle was 90\ensuremath{^\circ} and an electron flood gun
was used to compensate for sample charging. The spectrometer
operated at a base pressure of $1\cdot10^{-10}$ Torr. Evaporated
gold films supported on mica served as substrates. Each powdered
microcrystalline sample was dispersed in distilled-deionized water,
stirred for 5 minutes, and a few drops of the resulting suspension
were left to dry in air on a substrate. Directly after drying, the
samples were introduced into ultra high vacuum and placed on a He
cooled cryostat equipped with a Lakeshore cryogenic temperature
controller to explore the 50-350 K temperature range. All binding
energies were referenced to the nitrogen signal (cyanide groups) at
398 eV.\cite{ver06} No X-ray induced sample degradation was
detected. Spectral analysis included a Tougaard background
subtraction\cite{tougaard05} and peak deconvolution employing
Gaussian line shapes using the WinSpec program developed at the LISE
laboratory, University of Namur, Belgium.

\emph{Raman scattering.} Inelastic light scattering experiments in
the spectral region 2000-2300 cm$^{-1}$ (the spectral region of the
C-N stretching vibration) were performed in a 180\ensuremath{^\circ}
backscattering configuration, using a triple grating micro-Raman
spectrometer (T64000-Jobin Yvon), consisting of a double grating
monochromator (acting as a spectral filter) and a polychromator
which disperses the scattered light onto a liquid N$_{2}$ cooled CCD
detector. The spectral resolution was better than 2 cm$^{-1}$ for
the spectral region considered. Sample preparation was identical to
that for XPS measurements and samples were placed in a liquid He
cooled optical flow-cryostat (Oxford Instruments), where the
temperature was stabilized with an accuracy of 0.1 K throughout the
whole temperature range (from 300 to 50 K). A fraction of the second
harmonic output of a Nd:YVO$_{4}$ laser (532.6 nm, Verdi-Coherent)
was used as an excitation source and focused on the samples using a
50x microscope objective (Olympus, N.A. 0.5). The power density on
the samples was of the order of 600 W cm$^{-2}$.

\section{Results and Discussion}

\subsection{Magnetic susceptibility measurements}
\begin{figure}[htb]
\centering
\includegraphics[width=\figwidth]{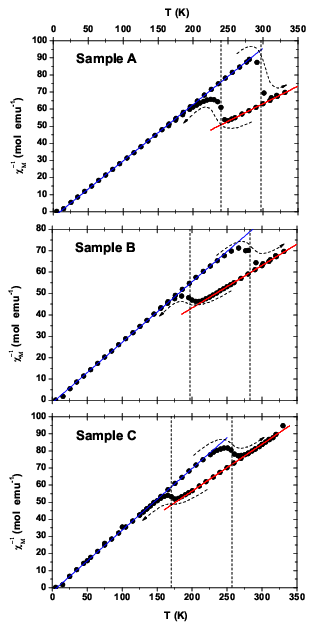}
\caption{$\chi^{-1}_{M}$ as a function of temperature for samples A
(\A), B (\B) and C (\C), showing the magnetic properties of the
materials. The Curie-Weiss law was fitted to the data of both the HT
and LT phases of the samples, the corresponding fits are indicated
by the blue (LT) and red (HT) lines in the graphs. From the fits,
experimental Curie constants ($C$) were extracted (see text). Dashed
arrows indicate the temperature dependence of the data as measured
in successive heating and cooling runs, while vertical indicate the
corresponding transition temperatures.} \label{Inverse_Chi}
\end{figure}

The inverse of the molar magnetic susceptibility, $\chi_{M}^{-1}$,
of the three samples is plotted in fig. \ref{Inverse_Chi}, as a
function of temperature. In all three samples the magnetic
properties show a thermal hysteresis; when heating the sample from 5
K up, a decrease in $\chi_{M}^{-1}$, accompanied by a decrease in
the slope of the $\chi_{M}^{-1},T$-curve, occurs at a characteristic
temperature T$_{1/2}$$\uparrow$, signaling the charge transfer (CT)
transition from the Fe$^{\textrm{II}}$-CN-Mn$^{\textrm{III}}$ (LT)
configuration to the Fe$^{\textrm{III}}$-CN-Mn$^{\textrm{II}}$ (HT)
configuration. Subsequent cooling shows the samples undergoing the
reverse transition at a temperature T$_{1/2}$$\downarrow$, which is
significantly lower than T$_{1/2}$$\uparrow$. Temperatures
T$_{1/2}$$\uparrow$ and T$_{1/2}$$\downarrow$ are defined as the
temperatures at which half of the CT transition has occurred in the
respective heating and cooling runs. Values of T$_{1/2}$$\downarrow$
= 240 K and T$_{1/2}$$\uparrow$ = 297 K for sample A (\A),
T$_{1/2}$$\downarrow$ = 197 K and T$_{1/2}$$\uparrow$ = 283 K for
sample B (\B) and T$_{1/2}$$\downarrow$ = 170 K and
T$_{1/2}$$\uparrow$ = 257 K for sample C (\C), as extracted from
corresponding $\chi_{M}T$-curves, yield hysteresis widths of 57, 86
and 87 K, respectively.

The magnetic properties of these samples are comparable to those
reported for other \general\
compounds.\cite{ver06,ohkoshi05,ohk02,cobo07} The samples discussed
here are consistent with the correlation between stoichiometry and
hysteresis properties as found by Ohkoshi \emph{et
al.}\cite{ohkoshi05} and Cobo \emph{et al.}\cite{cobo07} That is,
with increasing amount of Fe(CN)$_{6}$ vacancies,
T$_{1/2}$$\downarrow$, T$_{1/2}$$\uparrow$ and the amount of Rb in
the sample decrease, while
$\Delta$T=(T$_{1/2}$$\uparrow$-T$_{1/2}$$\downarrow$) and the amount
of H$_{2}$O in the sample increase. For all samples, the
susceptibility of both the LT and HT phase was fit to Curie-Weiss
behavior ($\chi_{M}^{-1}=\frac{(T-T_{c})}{C}$), the corresponding
fits are depicted by the blue (LT) and red (HT) lines in figure
\ref{Inverse_Chi}. Fits to the LT phase data were done in the
temperature range from 20 K up to approximately 10 K below the
respective T$_{1/2}$$\uparrow$ temperatures, while the fits to the
data of the HT phases were done in the range from approximately 10 K
above the respective T$_{1/2}$$\downarrow$ temperatures up to the
maximum measurement temperature (325 K). From these fits we
extracted the corresponding Curie constants (\emph{C}), which are
reported in table \ref{Magn measurements}. Theoretical values of the
Curie constant, $C$, were calculated\cite{chiT} for the assumed HT
and LT phases of samples A (\A), B (\B) and C (\C) and are also
given in table \ref{Magn measurements}. The fits also yielded the
characteristic temperatures $\theta$ for all samples in both the HT
and LT phases. Upon incorporation of ferromagnetic interactions in
the system (present in the HT phase of sample C, where
Fe\superscript{III} and Cu\superscript{II} ions interact
ferromagnetically), one would expect to see a shift of the negative
$\theta$ to smaller values (antiferromagnetic interactions remain
dominant). Indeed, such a shift can be seen ($\theta$ $= -7.7$ K,
$-17.6$ K and $-6.2$ K for sample A, B and C, respectively.).
However, the $\theta$ values extracted from the HT fits are the
result of an extrapolation over approximately 200 K, which would
make the observed shift be within the expected error bars. The
$\theta$ values corresponding to the LT fits do not show an obvious
trend ($\theta$ $= 9.6$ K, $4.1$ K and $6.5$ K, respectively), which
can be explained by the fact that in the LT the majority of the Fe
ions have assumed the $S = 0$, Fe\superscript{II} configuration. The
result is that the ferromagnetic Mn-Mn interaction dominates the
extracted parameters.

\begin{table*}[htb]
\centering \caption{Curie constants (\emph{C}, emu K mol$^{-1}$) and
'inactive fractions' (IFs, see text) for all samples} \label{Magn
measurements}
\setlength{\extrarowheight}{3pt}
\begin{tabular*}{17.0 cm}{@{\extracolsep{\fill}}ccD{.}{.}{3.3}cc}
\hline \hline
Sample (phase)  &   \multicolumn{1}{c}{$C$ (exp.)\footnotemark[1]} & \multicolumn{1}{c}{$C$ (calc.)\cite{chiT}} & IF (bulk, \%)\footnotemark[2]$^{,}$\cite{inactivefraction} & IF (surface, \%)\footnotemark[2]$^{,}$\cite{inactivefraction}\\
\hline
A (HT-phase)    &  4.87 $\pm$ 0.13    & 4.75                   & 1 $\pm$ 1.9  & 72 $\pm$ 0.5\\
A (LT-phase)    &  3.04 $\pm$ 0.09    & 3.03\footnotemark[3]   & 1 $\pm$ 1.9  & 72 $\pm$ 0.5\\
B (HT-phase)    &  5.03 $\pm$ 0.06    & 4.73                   & 28 $\pm$ 1.1  & 65 $\pm$ 0.5\\
B (LT-phase)    &  3.53 $\pm$ 0.05    & 3.07\footnotemark[3]   & 28 $\pm$ 1.1   & 65 $\pm$ 0.5\\
C (HT-phase)    &  3.65 $\pm$ 0.13    & 3.82                   & 24 $\pm$ 0.8   & 78 $\pm$ 0.5\\
C (LT-phase)    &  2.78 $\pm$ 0.11    & 2.45\footnotemark[3]   & 24 $\pm$ 0.8  & 78 $\pm$ 0.5\\
\hline \hline
\end{tabular*}
\footnotetext\protect[1]\protect{Errors are mostly due to the
uncertainty in the weight of the samples. This error is the same for
both phases of a particular compound. The fitting error is of the
order of 0.02 emu K mol$^{-1}$}.\\
\footnotetext\protect[2]\protect{The inactive fraction is defined as
the fraction of the magnetic species not undergoing the CT
transition, even though CT is stoichiometrically possible. I.e., the
magnetic species that do not undergo CT due to Fe:Mn
nonstoichiometry are excluded.}\\
\footnotetext\protect[3]\protect{The calculation of the C value
requires assumptions on the degree of charge transfer in the
compound. These calculated values correspond to the case of maximum
stoichiometrically possible degree of conversion in each of the
samples. See text for details.}
\end{table*}

The theoretical $C$ values calculated for sample A are, within the
experimental accuracy, in good agreement with those obtained from
the Curie-Weiss fit. In combination with the fact that in the
calculations\cite{chiT} all Fe ions and a stoichiometric amount of
Mn ions were assumed to undergo the charge transfer (CT) transition
in going from the HT to the LT configuration, these $C$ values
indicate that nearly all metal ions in sample A (which is very close
to the 'perfect' 1:1:1 stoichiometry) undergo CT, resulting in a
near maximum change in the magnetic properties. The calculated $C$
value for the HT phase of sample B is slightly higher than the
experimental value by about 0.3 emu K mol$^{-1}$. This might
indicate that the sample contains trace amounts of magnetic
impurities, which would increase the experimental C-value. The Curie
constant calculated for the LT phase of sample B is also lower than
its experimental value. Aside from aforementioned considerations,
this difference can be explained by realizing that while the
calculations assume a maximum possible degree of CT in the samples,
this is not necessarily the case (Sample B deviates somewhat from
the 1:1:1 stoichiometry). Thus, the difference can be ascribed to a
fraction of the magnetic species not undergoing the CT transfer
transition, even though the stoichiometry of the material would
allow for it (i.e. excluding the magnetic species that do not
undergo CT due to nonstoichiometry of Fe and Mn). This fraction will
hereafter be referred to as the "inactive fraction" (IF) of the
material. The estimated inactive fraction of each sample is also
given in table \ref{Magn measurements} for both bulk (or more
accurately, for bulk + surface material, as extracted from the
magnetic measurements\cite{inactivefraction}) and surface material
(as extracted from XPS spectra\cite{inactivefraction} (vide infra)).

The Curie constants calculated for sample C are somewhat lower due
to the partial substitution of Mn ions by Cu$^{\textrm{II}}$
entities ($S$ = 1/2) in the sample (the Cu ions are assumed to
remain divalent at all temperatures, which is verified by the XPS
measurements (vide infra)), the experimentally found value for the
HT phase is consistent with calculations. In the calculation of the
Curie constant of the LT phase, all the Mn ions and the
corresponding number of Fe ions were assumed to have undergone the
CT transition. Whether or not a metal ion undergoes the CT
transition in these systems is known, however, to depend on the
local stoichiometry in the system.\cite{ver06,cobo07,ohkoshi05}
Therefore, the calculation, which effectively ignores the effect of
the Cu ions on the CT probability of the other metal ions, is only a
lower limit to the LT phase Curie constant, which corresponds to the
case of maximum magnetic change across the transition for the given
stoichiometry (\C) (i.e., assuming IF = 0 in the material). And
indeed, the $C$-value found for the LT phase of sample B is somewhat
larger than this calculated lower limit, as would be intuitively
expected upon the incorporation of Cu ions into the lattice. For
comparison, assuming a random distribution of the Cu ions on the Mn
positions, the expected Curie constant for the LT phase would be
3.48 emu K mol$^{-1}$ (IF = 75.2 \%) if only the Fe ions surrounded
by 6 Mn ions (Fe[-CN-Mn]$_{6}$ clusters) are assumed to undergo the
CT transition, 2.91 emu K mol$^{-1}$ (IF = 33.1 \%) when also the Fe
ions in a Mn$_{5}$Cu environment would transfer an electron and 2.50
emu K mol$^{-1}$ (IF = 3.5 \%) when in addition even the Fe ions in
Mn$_{4}$Cu$_{2}$ surroundings would undergo CT. The experimentally
found $C$ value (2.78 emu K mol$^{-1}$) indicates the magnetic
properties are quite robust upon Cu-substitution; an estimated 76 \%
of the maximum possible CT transfer is still achieved (IF $\simeq$
24 \%), which is even more than is the case for sample B (IF =
$\simeq$ 28 \%). Moreover, it suggests the substitution of one (and
possibly even two) Mn ion by a Cu ion in the Fe[-(CN)-Mn]$_{6}$
cluster does not 'deactivate' the CT-capability of the Fe-center,
nor does it appear to reduce the cooperativity in the material,
since the CT transition occurs in a similar short temperature
interval in all samples (see fig. \ref{Inverse_Chi}).

\subsection{X-ray Photoemission Spectroscopy (XPS).}
XPS is a direct method for identifying the surface composition of a
compound as well as the oxidation state of the various elements at
the surface. Hence this technique is well suited to follow the phase
transition as a function of temperature. Indeed, as the structural
change of the compound is accompanied by a charge transfer between
metallic ions, XPS will accurately quantify the corresponding
changes in the oxidation states of the elements at the surface of
the material.

\begin{figure}[htb]
\centering
\includegraphics[width=5.0cm]{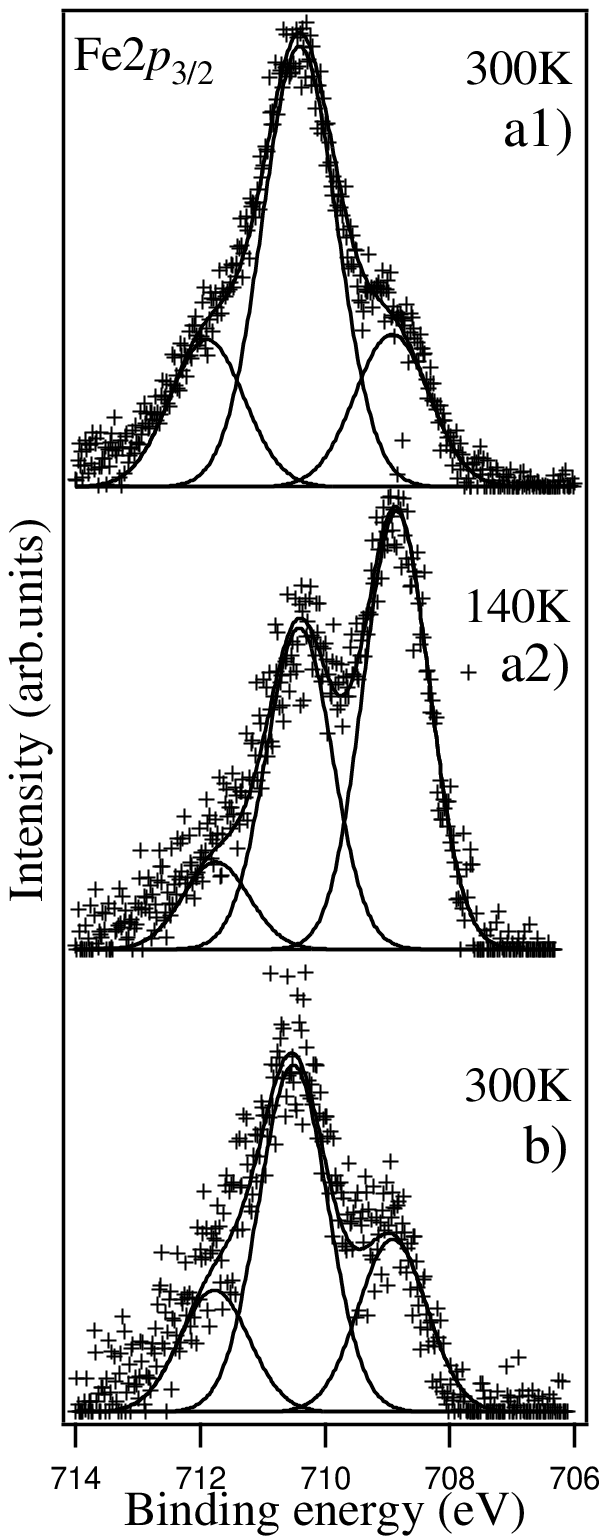}
\caption{Fe $2p$\subscript{3/2} core level photoemission spectra of
sample A (\A) collected at 300 and 140 K; fits to the raw data are
depicted as solid lines. Spectra labeled a) are recorded during the
cooling cycle, whereas the spectrum labeled b) refers to the
measurement done after warming back up to 300 K. The binding energy
scale is corrected for the temperature dependent shift (see text)
for clarity.} \label{XPS_A}
\end{figure}

\textbf{Sample A} (\A). Figure \ref{XPS_A} shows the Fe
$2p$\subscript{3/2} core level photoemission spectrum of sample A at
room temperature and at 140 K, as well as fits to the raw data. At
both temperatures, the Fe $2p$\subscript{3/2} signal consists of
three distinct contributions: the Fe\superscript{II} line at 708.8
eV binding energy, the Fe\superscript{III} line at 710.5 eV and the
Fe\superscript{III} satellite at 711.7 eV, the latter appearing 1.2
eV higher in binding than the Fe\superscript{III} main line, having
30 \% of its intensity.\cite{ver06} By comparing the relative
intensities of the Fe\superscript{II} and Fe\superscript{III}
components in the room temperature spectrum, one deduces that the
surface material of the compound is composed of 76 \%
Fe\superscript{III} and 24 \% Fe\superscript{II}. No spectral
changes are detected with respect to the room temperature data when
slowly cooling the sample to 252 K (rate $\sim$2 K/min., spectrum
not shown here), just above the HT to LT phase transition (see
figure \ref{Inverse_Chi}). As previously mentioned, further cooling
to 140 K induces a transition from the HT to the LT phase,
accompanied by an electron transfer from the Mn\superscript{II} to
the Fe\superscript{III} ions which is described as
Fe\superscript{III}($t^{5}_{2g}$,
S=$\nicefrac{1}{2}$)-CN-Mn\superscript{II}($t^{3}_{2g}e^{2}_{g}$,
S=$\nicefrac{5}{2}$) \textbf{$\rightarrow$}
Fe\superscript{II}($t^{6}_{2g}$,
S=$0$)-CN-Mn\superscript{III}($t^{3}_{2g}e^{1}_{g}$, S=$2$). The
photoemission spectrum of the Fe $2p$\subscript{3/2} clearly
qualitatively supports this transition, since the
Fe\superscript{III}/Fe\superscript{II} ratio after the transition is
48 \%/52 \%. However, the relative spectral intensities of the
Fe\superscript{III} and Fe\superscript{II} components also show that
the Fe\superscript{III}\textbf{$\rightarrow$}Fe\superscript{II}
conversion at the surface is far from complete (IF = 72 \%), in
contrast to the magnetic susceptibility measurements, which indicate
a near complete conversion for the bulk (see also table \ref{Magn
measurements}). This is explained by the surface sensitivity of the
XPS technique, since the surface composition and structure can
differ substantially from that of the corresponding bulk due to
surface reconstruction. Thus, probing this surface stoichiometry
using XPS shows the estimated\cite{inactivefraction} inactive
fraction of the surface material to be substantially larger in
sample A.

\begin{figure}[htb]
\centering
\includegraphics[width=\figwidth]{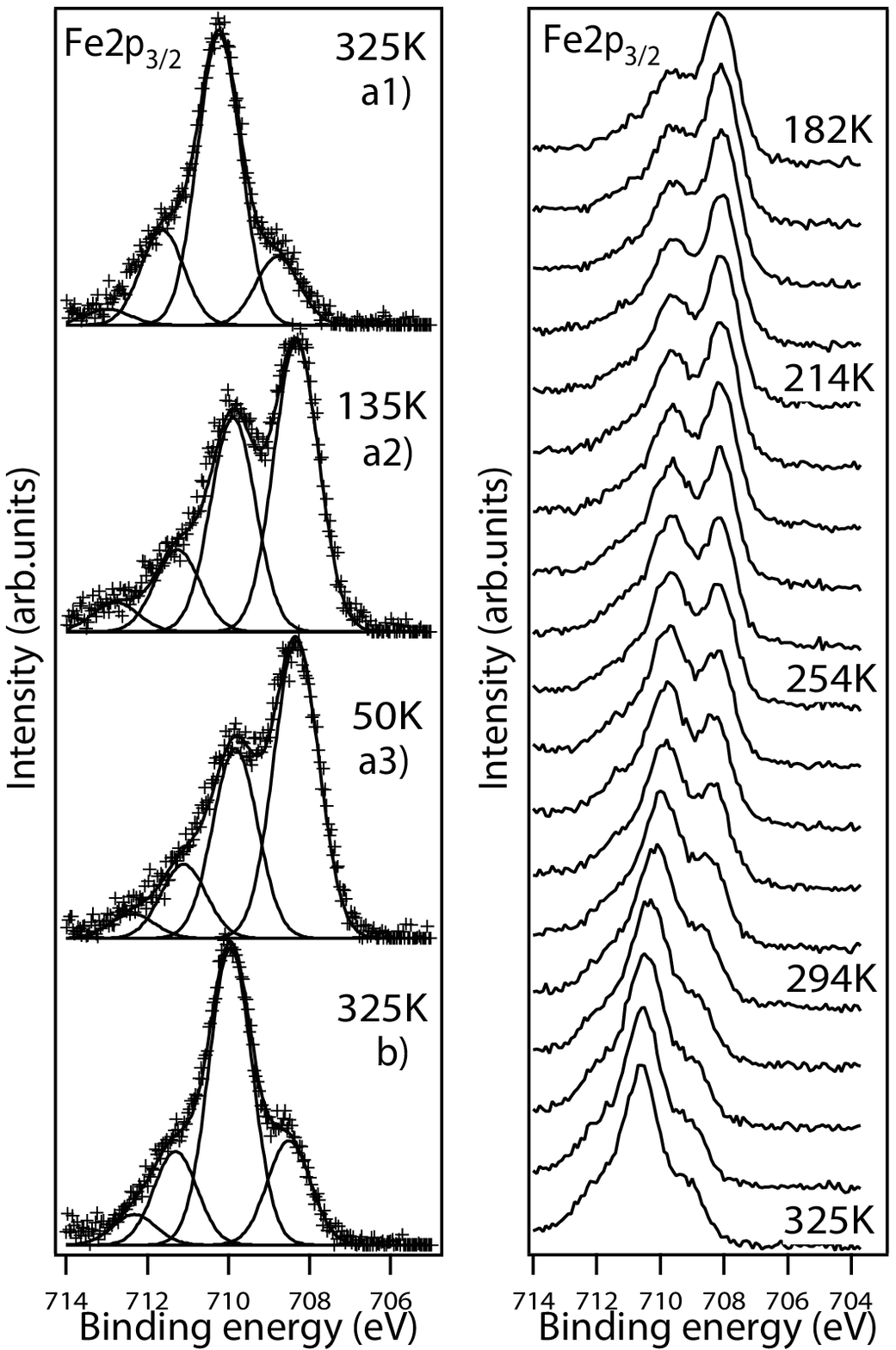}
\caption{Left panel: Fe $2p$\subscript{3/2} core level photoemission
spectra of sample B (\B) collected at 325, 135 and 50 K; solid lines
depict fits to the raw data. Spectra labeled a) are recorded during
the cooling cycle, whereas the spectrum labeled b) refers to the
measurement done after warming back up to 325 K. The binding energy
scale is corrected for the temperature dependent shift (see text)
for clarity. The right panel shows the evolution of the Fe
$2p$\subscript{3/2} core level photoemission spectrum as sample B is
warmed up from 182 to 325 K. Consecutive spectra are $\sim$8 K
apart. The binding energy scale is not rescaled here.} \label{XPS_B}
\end{figure}

\textbf{Sample B} (\B). The left panel of figure \ref{XPS_B} shows
the Fe $2p$\subscript{3/2} core level photoemission spectrum (fits
and raw data) of sample B for various temperatures, collected
starting from 325 K, then while cooling to a minimum temperature of
50 K and successively warming up back to 325 K (cooling and warming
rates both $\sim$2 K/min.). As in the photoemission spectrum of
sample A, the Fe $2p$\subscript{3/2} signal consists of three
distinct features: the Fe\superscript{II} line at 708.8 eV binding
energy, the Fe\superscript{III} line at 710.5 eV and the
Fe\superscript{III} satellite at 711.7 eV. Additionally, a fourth
small feature appears at the high binding energy side of the Fe
$2p$\subscript{3/2} signal. This contribution, which grows larger
with decreasing temperature, is attributed to a contribution of the
Fe\superscript{II} shake up satellite. By comparing the relative
Fe\superscript{III}/Fe\superscript{II} intensity ratio for 325 K,
135 K and 50 K one observes the same trend as mentioned for sample
A. The sample initially (at 325 K) consists of 85 \% of
Fe\superscript{III} and 15 \% of Fe\superscript{II} and upon
cooling, a decrease in the Fe\superscript{III} peak intensity is
observed, whereas the Fe\superscript{II} peak intensity
simultaneously increases, consistent with the charge transfer
between Fe\superscript{III} and Mn\superscript{II} ions. Spectra
recorded at 135 K and 50 K both reveal similar
Fe\superscript{III}/Fe\superscript{II} ratios, namely 51 \%/49 \%
and 50 \%/50 \%, respectively, in qualitative agreement with the
magnetic susceptibility measurements (figure \ref{Inverse_Chi}). As
for sample A, quantitative differences are due to the surface
sensitivity of the XPS technique and the increased inactive fraction
of the surface material with respect to that of the bulk ($\sim$72
\% vs. $\sim$28 \%, respectively).

\begin{figure}[htb]
\centering
\includegraphics[width=\figwidth]{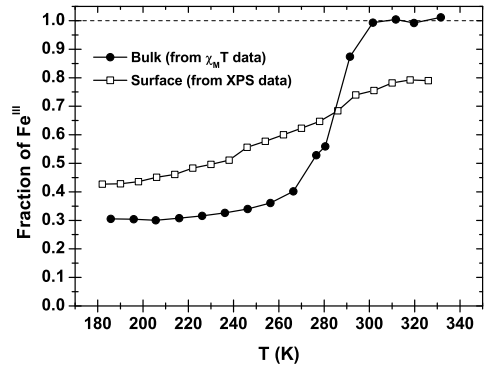}
\caption{Comparison between Fe\superscript{III} fractions in bulk
and surface material of sample B when heated through the CT
transition. Bulk data (estimated from $\chi_{M}$T values) shows a
much larger conversion fraction across a rather abrupt transition,
whereas surface data (as extracted from XPS spectra) shows a
substantially smaller and smooth conversion.} \label{comparison}
\end{figure}

In order to visualize the continuous conversion across the CT
transition in the surface material, we also recorded a series of Fe
$2p$\subscript{3/2} spectra while slowly warming up the sample at a
constant rate ($\sim$3.5 K/min.). The right panel of figure
\ref{XPS_B} shows this sequence starting at 182 K and reaching 325 K
with a temperature step of $\sim$8 K between spectra. The series
clearly shows a continuous conversion from the LT to the HT
configuration, testified to by the steadily decreasing intensity of
the Fe\superscript{II} peak and the increasing Fe\superscript{III}
spectral intensity. Thus, in contrast to the bulk material, the
surface material undergoes a smooth transition from the HT to the LT
phase, becoming progressively Fe\superscript{III}Mn\superscript{II}
due to the charge transfer from Fe\superscript{II} to
Mn\superscript{III} ions (see also fig. \ref{comparison}). In the
right panel of figure \ref{XPS_B} a slight shift of the Fe
$2p$\subscript{3/2} peak toward higher binding energy is observed
when increasing the temperature. A similar binding energy shift is
observed for the Mn spectra (not shown). Comparable shifts where
observed for all samples and cannot simply be attributed to sample
charging since the Rb and Fe shifts occur in opposite directions.
These shifts are due to a charge delocalization in the CN
vicinities.\cite{arrio05}

To illustrate the large differences between surface and bulk
behavior across the CT transition, the fraction of
Fe\superscript{III} versus temperature is plotted in figure
\ref{comparison} for both bulk and surface material of sample B,
during a heating process. Data for the bulk curve are estimated from
corresponding $\chi_{M}$T values, using the same formula as was used
to calculate Curie constants.\cite{chiT} Data for the surface curve
are extracted from the XPS spectra of the warming sequence in the
right panel of figure \ref{XPS_B}. The figure nicely visualizes the
differences: bulk material shows a high degree of conversion (IF
$\sim$ 28 \%) and displays a rather abrupt transition, while surface
material shows a much smoother transition of a substantially lower
amount of Fe ions. Also, neither the HT phase nor the LT phase of
the surface material consists of only one configuration
(Fe\superscript{III}Mn\superscript{II} or
Fe\superscript{II}Mn\superscript{III}). The differences are
attributed to a strongly increased degree of disorder and
inhomogeneity at the surface of the material, which increases the
inactive fraction and effectively eliminates cooperativity between
the metal centers, resulting in a smooth transition across a broad
temperature range.

\begin{figure}[htb] \centering
\includegraphics[width=\figwidth]{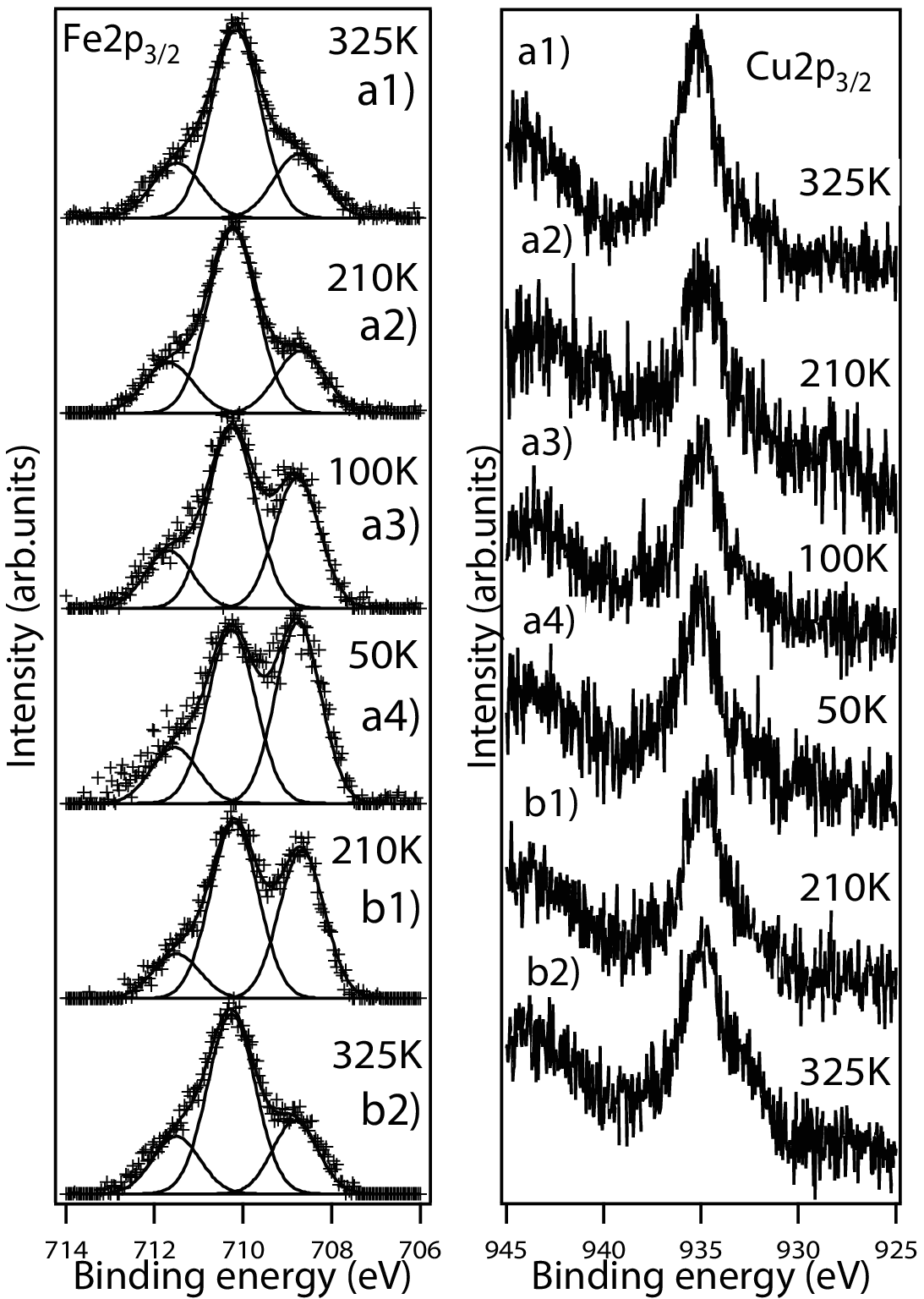}
\caption{Left panel: Fe $2p$\subscript{3/2} core level photoemission
spectra of sample C (\C) collected at 325, 210, 100 and 50 K. Solid
lines depict fits to the raw data. Spectra indicated with an a)
refer to measurements recorded during the cooling run, whereas
spectra indicated with a b) were recorded during the subsequent
heating run. The binding energy scale is corrected for the
temperature dependent shift (see text) for clarity. Right panel:
Corresponding Cu $2p$\subscript{3/2} core level photoemission
spectra of sample C recorded at different temperatures during the
same cooling and subsequent heating cycle.} \label{XPS_C}
\end{figure}

\textbf{Sample C} (\C). The left panel of figure \ref{XPS_C} shows
the Fe $2p$\subscript{3/2} core level photoemission spectrum (fits
and raw data) of sample C for six temperatures, recorded during
cooling down from 325 K to 50 K and subsequent warming up back to
325 K. As in samples A and B discussed above, the Fe
$2p$\subscript{3/2} signal of sample C consists of the same three
distinct features; the Fe\superscript{II} line at 708.8 eV, the
Fe\superscript{III} line at 710.5 eV and the Fe\superscript{III}
satellite at 711.7 eV. What distinguishes sample C from the others
is the fact that it contains $\approx$ 22 \% of copper on the Mn
positions. As expected, due to the inclusion of Cu into the lattice,
sample C (fig. \ref{XPS_C}) shows a lower absolute
Fe\superscript{III} to Fe\superscript{II} surface conversion in
comparison to samples A and B: starting out from 77 \%
Fe\superscript{III} and 23 \% Fe\superscript{II} at 325 K, the
sample shows a minimum ratio of 57 \% Fe\superscript{III} and 43 \%
Fe\superscript{II} at 50 K. Comparison of the surface inactive
fractions (in which the maximum degree of CT that is
stoichiometrically possible is taken into account) however, shows
the degree of conversion not to be anomalous (see table \ref{Magn
measurements}). As for the previous two samples A and B, the surface
IF is much higher than the corresponding bulk IF ($\sim$ 78 \% vs.
$\sim$ 24 \%), showing that also for sample C the phase transition
is far from complete at the surface of the sample.

Additional information about the role of Cu comes from the Cu
$2p$\subscript{3/2} core level photoemission data shown in the right
panel of figure \ref{XPS_C}. One observes that the main peak at 935
eV\cite{chawla92} remains constant in intensity and binding energy
throughout the temperature loop, revealing that the chemical
environment of the copper does not change across the phase
transition. One can therefore conclude that copper is not involved
in the charge transfer process.

\subsection{Raman Spectroscopy.}
Inelastic light scattering is employed in order to indirectly
determine the electronic and magnetic properties of the materials by
addressing the vibrational stretching mode of its CN-moieties. In
fact, the frequency of this vibrational mode, \cn, is highly
sensitive to the local environment of the CN-moiety. Upon
coordination, \cn\ will typically shift from its unbound ion
frequency, 2080 cm$^{-1}$, to a higher frequency, characteristic of
the local environment.\cite{nak86} The extent of this shift is
dependent on the electronegativity, valence and coordination number
of the metal ion(s) coordinated to the CN-group and whether they are
coordinated to the C or the N atom. Table \ref{CN-vibrations} shows
the typical frequency ranges where different CN stretching modes are
expected to be observed, when the cyano-moiety is in a bimetallic
bridge (an Fe-CN-M environment, where M = 3d metal ion). The given
frequency ranges are estimates based on literature and experimental
data of materials containing the specific or closely related
CN-environments (varying the N-bound metal
ion).\cite{nak86,reg90,sat99,ber88,ohk05,ver06} Across the thermal
phase transition, in addition to the intervalence charge transfer,
the \general\ lattice also contracts by approximately 10 \%. This
volume change decreases the average bond lengths in the system,
thereby generally increasing the vibrational frequencies.

\begin{table}[htb]
\centering \caption{Specific frequencies and frequency ranges for CN
stretching modes, \cn\, of CN-moieties in different environments in
Prussian Blue Analogues (M = 3d metal ion).} \label{CN-vibrations}
\setlength{\extrarowheight}{3pt}
\begin{tabular*}{8.0 cm}{@{\extracolsep{\fill}}cc}
\hline \hline
CN-moiety  &   \cn\ (cm$^{-1}$)\\
\hline
CN$^{-}$(aq)                       & 2080       \cite{nak86}\\
Fe$^{\textrm{II}}$-CN-Mn$^{\textrm{II}}$   & 2065 \cite{ber88}\\
Fe$^{\textrm{II}}$-CN-Mn$^{\textrm{III}}$   & 2095, 2096, 2114 \cite{ohk05,cobo07}\\
Fe$^{\textrm{II}}$-CN-Cu$^{\textrm{II}}$   & 2100 \cite{ber88}\\
Fe$^{\textrm{II}}$-CN-Mn$^{\textrm{III}}$   & 2114 \cite{cobo07}\\
\multirow{2}{*}{Fe$^{\textrm{III}}$-CN-Mn$^{\textrm{II}}$}  & 2146, 2152, 2155,\\
 & 2159, 2165, 2170 \cite{ber88,ohk05,ver06,cobo07}\\
Fe$^{\textrm{III}}$-CN-Cu$^{\textrm{II}}$   & 2172 \cite{ber88,ver06}\\
\hline
Fe$^{\textrm{II}}$-CN-M$^{\textrm{II}}$   & 2065-2100  \cite{nak86,reg90,sat99,ber88}\\
Fe$^{\textrm{II}}$-CN-M$^{\textrm{III}}$  & 2090-2140  \cite{nak86,ver06,sat99,ohk05}\\
Fe$^{\textrm{III}}$-CN-M$^{\textrm{II}}$  & 2146-2185  \cite{nak86,ver06,reg90,ber88,sat99,ohk05}\\
Fe$^{\textrm{III}}$-CN-M$^{\textrm{III}}$ & 2180-2210  \cite{nak86,reg90}\\
\hline
\end{tabular*}

\end{table}

\begin{figure}[htb]
\centering
\includegraphics[width=\figwidth]{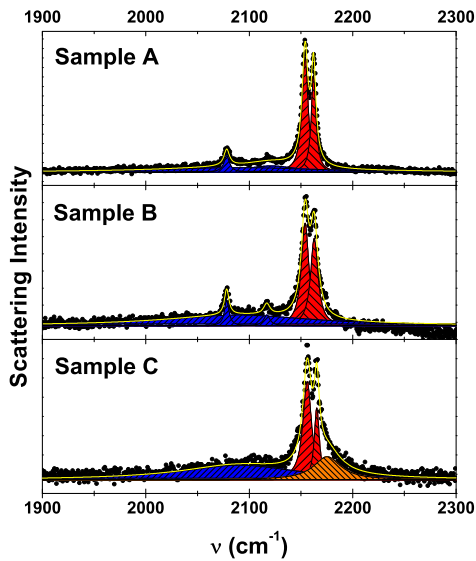}
\caption{Raman scattering spectra for all samples recorded at room
temperature. Multiple lorenztian contributions (color filled peaks)
were summed to obtain a fit (solid yellow line) to the data (black
circles). Red peaks correspond to the HT configuration of the
material (Fe$^{\textrm{III}}$-CN-Mn$^{\textrm{II}}$), while blue
peaks represent the LT configuration
(Fe$^{\textrm{II}}$-CN-Mn$^{\textrm{III}}$). The orange peak in the
lowest panel (sample C) represents the CN-vibrations corresponding
to Fe$^{\textrm{III}}$-CN-Cu$^{\textrm{II}}$ configurations.}
\label{HTspectra}
\end{figure}

The Raman spectrum of all three samples at room temperature (in the
HT phase) in the spectral window 2000-2250 cm$^{-1}$ is shown in
Fig. \ref{HTspectra}. Samples were heated to 330 K prior to
measurements to ensure the samples were in their HT phase. Group
theory analysis predicts that the vibrational stretching mode of the
free CN$^{-}$ ion ($A_{1}$ symmetry) splits up into an $A_{1}$, an
$E$ and a $T_{2}$ normal mode, when the CN moiety is placed on the
$C_{2v}$ site of the $F$\={4}$3m$ ($T_{d}^{2}$) space group (the
space group in the HT phase\cite{mor02,mor03,kat03,ver06}).

From the corresponding Raman tensors\cite{kuz98} it is clear that
the $A_{1}$, and $E$ normal modes are expected to be observed in the
parallel polarization spectra of figure \ref{HTspectra} (due to
their non-zero diagonal tensor components). Indeed, all spectra show
a double peak structure (red colored lines), with vibrations at 2156
and 2165 cm$^{-1}$, which are ascribed to the $A_{1}$ and $E$ normal
modes of the Fe$^{\textrm{III}}$-CN-Mn$^{\textrm{II}}$ (HT phase)
moieties. These peaks are observed in the expected frequency range
(see table \ref{CN-vibrations}) and are consistent with
IR-data\cite{ohk05,ber88,sat99} and previous Raman
measurements.\cite{cobo07,ver06,ver08} Also, the spectra of all
samples show some additional Raman intensity at lower wavenumbers
(blue colored lines), consistent with the presence of
LT-configuration moieties
(Fe$^{\textrm{II}}$-CN-Mn$^{\textrm{III}}$, table
\ref{CN-vibrations}), as was observed in the XPS data. In samples A
and B, one can distinguish clear peak features at $\sim$2080 and
$\sim$2118 cm$^{-1}$ on top of the broad LT-phase intensity, whereas
sample C shows only a featureless broad band. The presence of
Fe$^{\textrm{III}}$-CN-Cu$^{\textrm{II}}$ cyano bridges in sample C
is reflected in its spectrum through a shoulder on the high
wavenumber side of the double peak structure (orange filled peak).
The incorporation of Cu$^{\textrm{II}}$ ions into the lattice is
also evident in the width of the lines; inhomogeneous broadening as
a result of the increased degree of disorder causes the peaks in
sample C to be broader with respect to those in samples A and B,
which is arguably also the reason the LT peak features are not
distinguishable in sample C. Unfortunately, no reliable quantitative
estimation regarding the Fe\superscript{III}/Fe\superscript{II}
ratios can be made from the intensities of the Raman lines, since
these involve a phonon-dependent proportionality factor. In
addition, possible photo-induced LT to HT switching at these
temperatures would further complicate such estimations. Nonetheless,
changes in the intensity of one phonon as a function of, for
instance, temperature can give information on relative quantities of
the given phonon.

\subsubsection*{Temperature dependence}

Table \ref{CN-vibrations} illustrates the effect of the variation of
valence of either of the metal ions in an Fe-CN-M (cyano) bridge on
the vibrational stretching frequency of the CN-moiety (\cn). The
predominant trend is an increase in \cn\ with increasing oxidation
state of either of the metal ions, where the valence of the C-bound
metal ion appears to have a larger effect than that of the N-bound
metal ion.\cite{nak86,mil99} Thus, across the temperature induced CT
transition in these materials, when the local environment changes
from Fe$^{\textrm{III}}$-CN-Mn$^{\textrm{II}}$ to
Fe$^{\textrm{II}}$-CN-Mn$^{\textrm{III}}$ in the cooling run, a net
downshift in vibrational frequencies is expected.

\begin{figure*}[htb]
\centering
\includegraphics[width=17.5 cm]{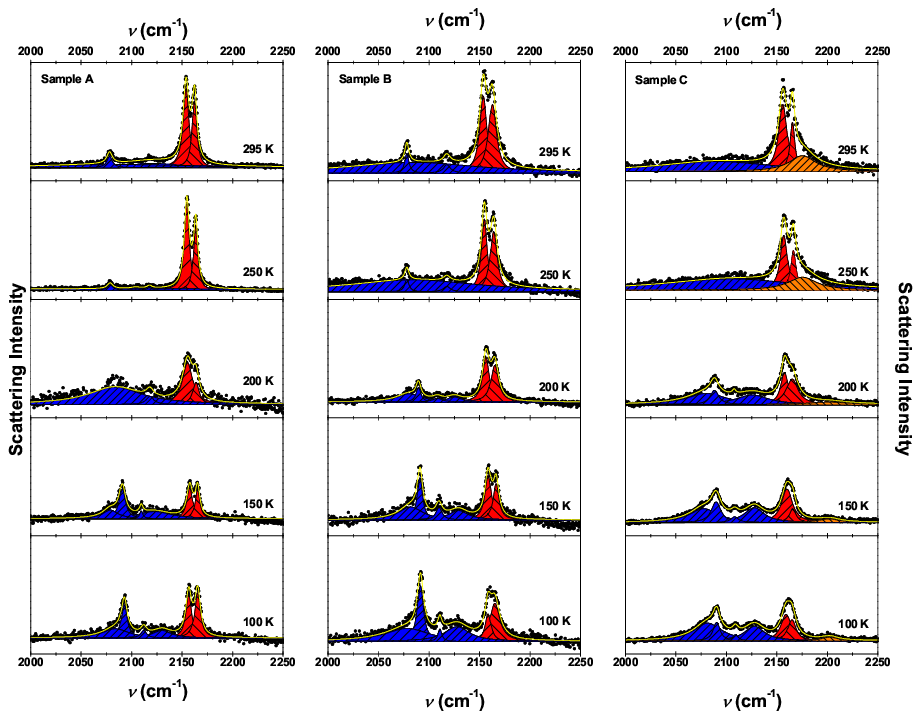}
\caption{Temperature dependence of the Raman spectrum of samples A
(\A), B (\B) and C (\C) across the hysteresis range in a cooling
run. Multiple lorenztian contributions (color filled peaks, red =
HT, blue = LT, orange = Fe$^{\textrm{III}}$-CN-Cu$^{\textrm{II}}$)
were summed to obtain a fit (solid yellow line) to the data (black
circles). Spectra are normalized to the total integrated scattering
intensity in the 2000-2250 cm$^{-1}$ window.} \label{Tdep}
\end{figure*}

The evolution of the parallel polarization Raman spectrum of the
three samples across the hysteresis range in a cooling run is shown
in figure \ref{Tdep}. At a temperature of 250 K, all samples still
show the spectrum typical of the HT phase, which is discussed above.
Upon cooling the samples through their respective CT transitions,
however, the intensity of the HT (red) lines goes down and
simultaneously, the intensity of the broad signal at lower
wavenumbers increases and evolves into several new peaks at
$\sim$2202, $\sim$2125, $\sim$2108, $\sim$2089 and $\sim$2080
cm$^{-1}$ (blue lines). In addition, the HT lines are slightly
shifted to higher frequencies, due to the contraction of the
lattice. This suggests that, although no CT has occurred in the
inactive fraction of the samples, the lattice does contract. Even
though no quantitative Fe\superscript{III}/Fe\superscript{II} ratio
can be extracted from the Raman spectra, the presence of both HT
(red) and LT (blue) lines seem to indicate an incomplete CT
transition, in accordance with the XPS results. Consequently,
multiple different Fe-CN-M environments are present in the LT phase
of the samples, which explains the large number of lines observed in
their Raman spectra. Next to the residual HT
Fe$^{\textrm{III}}$-CN-Mn$^{\textrm{II}}$ configuration (red lines)
and the LT Fe$^{\textrm{II}}$-CN-Mn$^{\textrm{III}}$ configuration,
also Fe$^{\textrm{II}}$-CN-Mn$^{\textrm{II}}$ and
Fe$^{\textrm{III}}$-CN-Mn$^{\textrm{III}}$ configurations are
present in the LT phase. The latter configuration is confidently
assigned to the 2202 cm$^{-1}$ line, since it is the only
configuration in which \cn\ is expected to increase (Table
\ref{CN-vibrations}). Less straightforward is the assignment of the
other configurations since their cyano vibrations are expected to
occur in overlapping frequency ranges. In addition, the modes
arising from these configurations are also expected to have split
into multiple normal modes due to the crystal symmetry. For
comparison, when assuming the space group $I$\={4}$m2$ (that of
\general\ in the LT phase\cite{ver06,mor02,kat03}), the CN
vibrational mode splits up into 2 A$_{1}$, a B$_{1}$, a B$_{2}$ and
an E mode, of which three (2 A$_{1}$ and B$_{1}$) would be observed
in parallel polarization spectra. Nevertheless, based on the
expected frequencies (table \ref{CN-vibrations}) and the expected
symmetry splitting the 2125, 2108 and 2089 cm$^{-1}$ lines are
assigned to symmetry split normal modes of the
Fe$^{\textrm{II}}$-CN-Mn$^{\textrm{III}}$ LT configuration, in
agreement with Cobo \textit{et al.}\cite{cobo07}, while the 2080
cm$^{-1}$ line is again tentatively assigned to the
Fe$^{\textrm{II}}$-CN-Mn$^{\textrm{II}}$ configuration. Due to the
fact that the Fe$^{\textrm{II}}$-CN-Mn$^{\textrm{II}}$ and
Fe$^{\textrm{III}}$-CN-Mn$^{\textrm{III}}$ configurations occur at
the interface between a metal ion that has and a metal ion that has
not undergone CT (generally in an environment with local
inhomogeneities), the symmetry splitting of the modes assigned to
these configurations is lost in the inhomogeneously broadened width
of their corresponding lines (2080 and 2202 cm$^{-1}$). One may also
expect the Raman spectrum of sample C in the LT phase to show
vibrations arising from the Fe$^{\textrm{II}}$-CN-Cu$^{\textrm{II}}$
($\approx$ 2100 cm$^{-1}$) and
Fe$^{\textrm{III}}$-CN-Cu$^{\textrm{II}}$ ($\approx$ 2175 cm$^{-1}$)
configurations, however, due to their low intensity and frequency
overlap with more intense lines, it is not possible to resolve these
lines in the present data. Overall, the temperature dependence of
the Raman spectrum is in general good agreement with the XPS
measurements, nicely demonstrating the temperature-induced CT
transition. In addition to the changes in the vibrational spectra,
all samples show a pronounced color change across the CT transition.
The samples are substantially darker in their LT phase and are more
susceptible to laser-induced degradation.

\subsubsection*{Photoactivity}

\begin{figure}[htb]
\centering
\includegraphics[width=\figwidth]{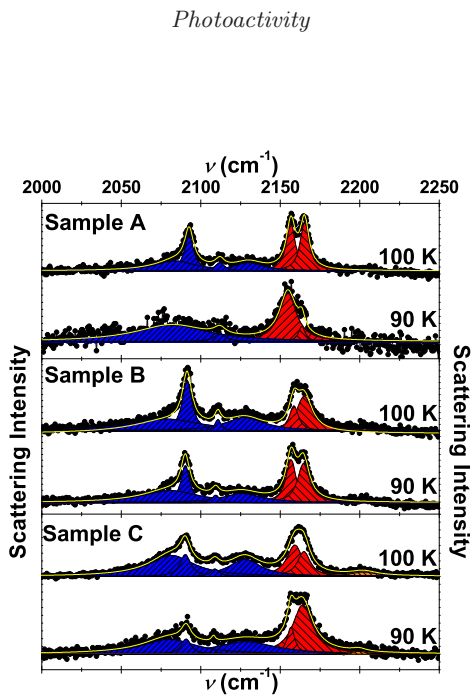}
\caption{Raman spectra of all three samples at 100 K and 90 K below
the hysteresis in a cooling run. In all samples, the spectra at 90 K
show an increase in intensity of the HT lines and a decrease of the
LT lines with respect  to the corresponding 100 K spectra. This is
attributed to the photoactivity of the material (see text). All
spectra are normalized to the total integrated scattering intensity
in the 2000-2250 cm$^{-1}$ spectral window} \label{90K}
\end{figure}

Remarkably, at a temperature of 90 K, the Raman spectra of the
samples show the material has regained some intensity in the HT
(red) lines at the cost of the LT (blue) lines (see fig \ref{90K}),
breaking the general trend of an increasing LT fraction with
decreasing temperature. Also, the double peak structure in sample C
again shows a clear shoulder on the high frequency side, arising
from the Fe$^{\textrm{III}}$-CN-Cu$^{\textrm{II}}$ cyano bridge.
These features are explained in terms of the photoactivity of the
material upon excitation of the sample using the Raman laser probe
(532 nm). Around 90 K, \general\ has been shown to be photo-excited
into a metastable state upon 532 nm laser
excitation.\cite{mo03,cobo07} The resulting metastable phase is
described as 'HT-like', meaning that the predominant valence
configuration is Fe$^{\textrm{III}}$-CN-Mn$^{\textrm{II}}$, which is
consistent with the present and earlier\cite{cobo07} Raman spectra
at these temperatures. A local laser induced heating effect is
excluded, since the effect is not observed just below the
corresponding hysteresis loops. Also, the metastable phase is stable
in absence of laser irradiation and only relaxes to
Fe$^{\textrm{II}}$-CN-Mn$^{\textrm{III}}$ ground state above a
certain relaxation temperature (see below). In addition to the
spectral changes, the sample is also observed to undergo a change in
its optical properties under 532 nm excitation at this temperature:
the excited material takes the substantially lighter HT appearance
(the inverse of the color change seen when cooling through the CT
transition). A similar effect is seen around 90 K: spectra recorded
at 100, 80, 70, and 50 K also show increased intensity in the HT
lines. The effect becomes increasingly less pronounced as the
temperature deviates more from 90 K, which appears to be the
temperature of maximum efficiency in photo-conversion. As the
photo-conversion is accompanied by a color change, the
(meta)stability of the photo-excited state is easily monitored
visually, observing the sample color under a microscope.
Consequently, the photo-excited 'HT-like' state was found to be
persistent after excitation for at least 2 hours (without laser
irradiation) at 90 K, showing no signs of relaxation to the darker
LT ground state. During a subsequent slow heating ($\sim$ 0.5
K/min.) process the photo-excited state was visually monitored and
found to relax to the LT ground state at a temperature of $\sim$ 123
K (see movie clip in Supporting Information), consistent with the
relaxation temperature reported by Tokoro \emph{et
al.}\cite{tokoro03} (120 K). A more elaborate study of the
photo-conversion of the material as a function of temperature,
excitation wavelength and intensity is required to elucidate the
nature of this fascinating metastable photo-induced phase, the
conversion mechanism involved and the striking temperature
dependence of the effect.

\section{Conclusions}
In conclusion, this work demonstrates the temperature-induced charge
transfer transition in different Prussian Blue Analogue samples
through a number of different experimental techniques, revealing the
substantially reduced conversion factor of the surface material with
respect to the bulk material. All three techniques, magnetic
susceptibility measurements, XPS and Raman spectroscopy, show the
thermally induced charge transfer transition, which can be described
as Fe\superscript{III}($t^{5}_{2g}$,
S=$\nicefrac{1}{2}$)-CN-Mn\superscript{II}($t^{3}_{2g}e^{2}_{g}$,
S=$\nicefrac{5}{2}$) \textbf{$\rightarrow$}
Fe\superscript{II}($t^{6}_{2g}$,
S=$0$)-CN-Mn\superscript{III}($t^{3}_{2g}e^{1}_{g}$, S=$2$).
Magnetic measurements indicate the bulk material shows a high degree
of conversion (near maximal) in sample A (\A), while the conversion
fraction is lower in sample B (\B). This is according expectation,
as sample A is much closer to a Rb:Mn:Fe stoichiometry of 1:1:1.
However, X-ray photoemission spectroscopy reveals a substantially
lower HT $\rightarrow$ LT conversion at the sample surface of all
samples, the fraction of metal centers not undergoing the charge
transfer transition is by far dominant at the surface, even in the
highly stoichiometric sample A. This shows the \textit{intrinsic}
incompleteness of such systems to be due to surface reconstruction.
Additionally, the CT transition is found to be much more smooth and
continuous at the surface of the samples, due to the fact that
cooperativity is effectively eliminated when the HT to LT conversion
fraction is very low.

 Though substitution of a fraction of the
Mn\superscript{II} ions by Cu\superscript{II} ions (in sample C, \C)
is shown to reduce the degree of LT to HT conversion, the reduction
is comparable to the fraction of Cu ions being substituted; for
sample C (\C), which contains 22 \% of Cu on the Mn-positions, still
76 \% of the maximum possible Fe\superscript{III}($t^{5}_{2g}$,
$S=\nicefrac{1}{2}$)-CN-Mn\superscript{II}($t^{3}_{2g}e^{2}_{g}$,
$S=\nicefrac{5}{2}$) \textbf{$\rightarrow$} Fe\superscript{II}
($t^{6}_{2g}$, $S=0$)-CN- Mn\superscript{III}($t^{3}_{2g}e^{1}_{g}$,
$S=2$) conversion is observed, which is comparable to the
percentages found in sample B, which has no Cu incorporated in the
lattice. Thus, the random substitution has little to no effect on
the charge transfer capability of individual metal clusters. In
fact, a simple numerical analysis shows local
Fe[-CN-Mn]\subscript{5}[-CN-Cu] and even
Fe[-CN-Mn]\subscript{4}[-CN-Cu]\subscript{2} clusters are not
deactivated regarding charge transfer.  Temperature dependent Raman
spectroscopy is in agreement with above results, clearly displaying
the charge transfer transition to be incomplete in all samples.
Summarizing, these results show that the maximum total degree of HT
$\rightarrow$ LT conversion in these systems, found for highly
stoichiometric samples is intrinsically limited by the fact that
surface reconstruction deactivates metal clusters at the surface of
the material regarding charge transfer.

    At temperatures of 50-100 K, a remarkable photoactivity of the
material is observed. Raman spectra in this temperature interval
show the material to be photo-excited from the LT state, into a
metastable "HT-like" state, meaning that the predominant valence
configuration in this state is
Fe\superscript{III}Mn\superscript{II}. This photo-conversion, which
appears to be most efficient at 90 K, is accompanied by substantial
color changes and is found to be stable below a relaxation
temperature of $\sim$ 123 K. How this state is related to the
photo-excited (meta)stable phase at very low temperatures is not
clear at the moment, further investigations are required to
determine its exact nature and fascinating temperature dependence.
\\

\textbf{Acknowledgment.} The authors would like to thank Roland
Hubel for technical support during the XPS measurements at the IWF
in Dresden. This work is part of the research programme of the
'Stichting voor Fundamenteel Onderzoek der Materie (FOM)', which is
financially supported by the 'Nederlandse Organisatie voor
Wetenschappelijk Onderzoek (NWO)'
\\


\begin{thebibliography}{42}
\bibitem{inactivefraction}
    The inactive fraction (IF) is defined as the fraction of
the magnetic species not undergoing the ET transfer transition, even
though ET is stoichiometrically possible. I.e., the magnetic species
that do not undergo ET due to Fe:Mn nonstoichiometry are excluded.
The estimated bulk IF values were calculated by matching the formula
$C=\frac{N_{A}\mu_{B}^{2}g^{2}}{3k_{B}}\sum_{i}c_{i} S_{i}(S_{i}+1)$
(see reference above) to the experimental $C$-value of the
corresponding LT phase. The estimated surface IF values are
calculated from the experimental
Fe\superscript{III}/Fe\superscript{II} ratios extracted from
corresponding XPS spectra. For sample C the IF is corrected for the
material that is unable to undergo ET due to the $\#$Fe$>$$\#$Mn
nonstoichiometry.
\end{thebibliography}
\end{document}